\documentclass[doublecol, figures]{epl2}
\pdfoutput=1
\usepackage{amsmath}
\usepackage[]{hyperref}
\begin{document}
\title{Energy dissipation of moved magnetic vortices}
\author{Martin P. Magiera \thanks{E-mail:
    \email{martin.magiera@uni-due.de}}}

\shortauthor{M. P. Magiera}
\institute{
Faculty of Physics and CeNIDE, University of Duisburg-Essen,
D-47048 Duisburg, Germany, EU
}
\date{\today}

\abstract{ A two-dimensional easy-plane ferromagnetic substrate,
  interacting with a dipolar tip which is magnetised perpendicular
  with respect to the easy plane is studied numerically by solving the
  Landau-Lifshitz Gilbert equation. The dipolar tip stabilises a
  vortex structure which is dragged through the system and dissipates
  energy. An analytical expression for the friction force in the $v
  {\rightarrow} 0$-limit based on the Thiele equation is
  presented. The limitations of this result which predicts a diverging
  friction force in the thermodynamic limit, are demonstrated by a
  study of the size dependence of the friction force.  While for small
  system sizes the dissipation depends logarithmically on the system
  size, it saturates at a specific velocity dependent value. This size
  can be regarded as an effective vortex size and it is shown how this
  effective vortex size agrees with the infinite extension of a vortex
  in the thermodynamic limit. A magnetic friction
    number is defined which represents a general criterion for the
    validity of the Thiele equation and quantifies the degree of
    nonlinearity in the response of a driven spin configuration.}
\maketitle

\section{Introduction}
Vortices in magnetic layers have been known for a long time
\cite{Belavin1975, Nikiforov83, Huber82, Gouvea1989, Shinjo2000,
  Wachowiak2002, Costa2011}. Although a vortex represents a strong
excitation with a high energy, it has a long life time for topological
reasons: Each vortex can be characterised by the vorticity, which is a
conserved quantity for the entire system.  To annihilate a vortex in a
closed system, an antivortex (which has a negative vorticity and also
cannot be created spontaneously) is required. In an open system,
(anti)vortices can be created at the system boundary. A second
quantity related to vortices is the polarisation, the out-of-plane
magnetisation of the vortex core, which may adopt two states.
Therefore a vortex state represents a bit, which can be easily probed
\textit{e.g.}\ with GMR sensors, as those used in reading heads of
magnetic hard disks.  It can be manipulated at very short timescales
(down to picoseconds) by magnetic field pulses \cite{Xiao2006,
  Hertel2007}, alternating magnetic fields \cite{Waeyenberge2006} or
spin-polarised currents \cite{Yamada2007}, making magnetic vortices
promising candidates for non-volatile storage concepts.  In a previous
work, it has been shown that vortex states may also be generated or
annihilated by a magnetic tip scanning a magnetic substrate, when the
interaction strength between tip and substrate as well as the scanning
velocity is appropriately adjusted \cite{Magiera12}. The manipulation
of vortices by the tip of a magnetic force microscope has been
observed experimentally in type-II superconductors
\cite{Auslaender2008}. Recently, also the switching of single
skyrmions (magnetic vortices with the tail magnetisation pointing
anti-parallel to the core magnetisation, \cite{Skyrme1962}) at thin
PdFe films has been realised \cite{Romming2013}. The switching is here
induced by a spin-polarised current, injected by a scanning-tunneling
microscopy tip which is positioned above the substrate.

In this work, the focus lies on the energy dissipation occurring when a
vortex is dragged through the ferromagnet, which leads to a friction
force decelerating the magnetic tip. This \textit{magnetic friction}
force is a direct consequence of the non-equilibrium nature. A
prototype for studies of magnetic friction is the Ising model,
subdivided in (at least) two subsystems, one of which is shifted with
respect to the other one with a rate representing a velocity
\cite{Kadau08, Hucht10, Hilhorst11, Angst11}. Magnetic friction in the
Potts model \cite{Igloi11} as well as in sheared geometries
\cite{Hucht12} has also been observed.

Experimental evidence of magnetic friction forces has been provided by
magnetic exchange force microscopy experiments recently
\cite{Wiesendanger2012, Heinrich2012}, where a magnetic tip dragged a
single magnetic atom across a magnetic surface. Such a system has been
modelled in our earlier works
\cite{Fusco08,Magiera09a,Magiera09b,Magiera10,Magiera11,Magiera11b}. The
microscopic mechanism leading to a friction force, which depends
linearly on the scanning velocity of the tip \cite{Magiera09a,
  Magiera11} was identified, as well as the influence of temperature
on the friction force \cite{Magiera09b, Magiera10}. An explicit
comparison and unification of results in the Ising model and results
in the Heisenberg model was provided in \cite{Magiera11b}. A
field-theoretical treatment is presented in \cite{Demery10}. In this
Letter the magnetic friction force for a system containing a magnetic
vortex is explicitly calculated.

\section{System}
The system consists of $L^2$ classical Heisenberg spins $\mathbf S_i =
\boldsymbol{\mu}_i/\mu_s$ on a square grid, where $\mu_s$ is a
material specific saturation magnetisation. The Hamiltonian contains
two terms, corresponding to a substrate and a tip part,
\begin{equation}
\mathcal H = \mathcal H_\mathrm{sub} + \mathcal H_\mathrm{tip}.
\end{equation}
To describe the substrate I use an isotropic exchange between nearest
neighbours with interaction constant $J$ and equivalent anisotropy in
$x$- and $y$-direction,
\begin{equation}
\mathcal H_\mathrm{sub} = -J \sum_{\left < i,j \right >} \mathbf S_i
\cdot \mathbf S_j - d_z \sum_i S_{i,z}^2,
\label{eq2}
\end{equation}
where $d_z{<}0$ leads to an easy plane (spins tend to align in the
$xy$-plane) and stabilises magnetic vortices. The value
$d_z={-}0.1J$ is used in this Letter. The explicit value of $d_z$ only
influences the vortex core radius, but not the phenomenology described
below.

The moved tip interacts with the substrate via a dipolar interaction,
\begin{equation}
 \mathcal H_\mathrm{tip} = - w \sum_i {\frac{3
  ~(\mathbf S_i \cdot \mathbf e_i) (\mathbf S_\mathrm{tip} \cdot
  \mathbf e_i) - \mathbf S_i \cdot \mathbf S_\mathrm{tip} } 
  {R_i^3}}, 
\label{eq3}
\end{equation}
where $R_i = \left |\mathbf R_i \right |$ denotes the norm of the
position of spin $i$ relative to the tip $\mathbf R_i = \mathbf r_i -
\mathbf r_\mathrm{tip}$, and $\mathbf e_i$ its unit vector $\mathbf
e_i = \mathbf R_i / R_i $. $\mathbf r_i$ and $\mathbf r_\mathrm{tip}$
are the position vectors of the substrate spins and the tip
respectively.  $w$ is a free parameter that quantifies the
dipole-dipole-coupling between the substrate spins and the tip, thus
controlling the strength of the tip. I use
$\mathbf{S}_\mathrm{tip}=(0,0,-1)$.  The tip is moved with constant
velocity $(v,0,0)$ two lattice constants above the substrate, along
the middle line between two spin rows. In a previous work,
ref.~\cite{Magiera12}, a regime of $w$- and $v$-values has been
identified, leading to stable vortices dragged through the
system. Results presented here are restricted to this regime. The
height $z$ at which the tip is dragged above the substrate as well as
the value $w$ and the fact that a dipolar interaction is used (and not
\textit{e.g.}\ a monopole approximation as discussed in
\cite{Haberle2012}) are of little relevance for the structure of the
vortices, as well as for the results presented below.

Open boundary conditions are used in $y$-direction. Co-moving
open boundaries are implemented in $x$-direction: When the
tip is moved by exactly one lattice constant (one \textit{cycle}), the
foremost spin row is duplicated, and the last one is deleted, see also
\cite{Magiera09a}. In this way the simulation can go on indefinitely.
The substrate spins follow the Landau-Lifshitz-Gilbert (LLG) equation
\cite{LandauLifshitz1935, Gilbert2004},
\begin{equation}
 \frac{\partial}{\partial t} \mathbf S_i = -\frac{\gamma}{(1 +
  \alpha^2) \mu_s} \left[\mathbf S_i \times \mathbf h_i + \alpha ~
  \mathbf S_i \times(\mathbf S_i \times \mathbf h_i) \right],
\end{equation}
with saturation magnetisation $\mu_s$, gyromagnetic ratio $\gamma$,
the phenomenological damping constant $\alpha$ (high damping values
$\alpha{=}0.5$ and $\alpha{=}0.3$ are used in this Letter to reach a
steady state in a ``short'' simulation time) and the local field $
\mathbf h_i = - \partial \mathcal H/\partial \mathbf S_i.  $ It
produces Larmor precession with frequency $\left | \mathbf h_i\right|
\gamma/\mu_s$, and a damping in the direction of the local field. One
may define a characteristic frequency $\omega = \gamma J/\mu_s$. To
solve the LLG equation the Heun integration scheme \cite{Garcia98} is
used.  To calculate the energy dissipation, the expression
\cite{Magiera09a}
\begin{equation}
  P_\mathrm{diss} = -  \sum_i \mathbf h_i \cdot \partial_t \mathbf S_i = \frac{\gamma}{\mu_s} \frac{\alpha}{1+\alpha^2} \sum_i \left ( \mathbf S_i \times \mathbf h_i \right )^2
\label{eq:Pdiss}
\end{equation}
is used, which leads to the correct magnetic friction force at zero
temperature,
\begin{equation}
  F = \frac{\left < P_\mathrm{diss} \right >}{v},
\end{equation}
after averaging over at least one cycle $a/v$ in the steady
state. Note that reaching of the steady state, especially for large
system sizes, requires long simulation runs (the data points presented
below are determined after $10^8$ integration steps).  In the
following natural units are used (time $\omega^{-1}=\mu_s/(\gamma J)$,
energy $J$ and length $a$).

\section{Friction in the quasi-static limit}
\begin{figure}[bt]
\centering
 \includegraphics[width=\columnwidth]{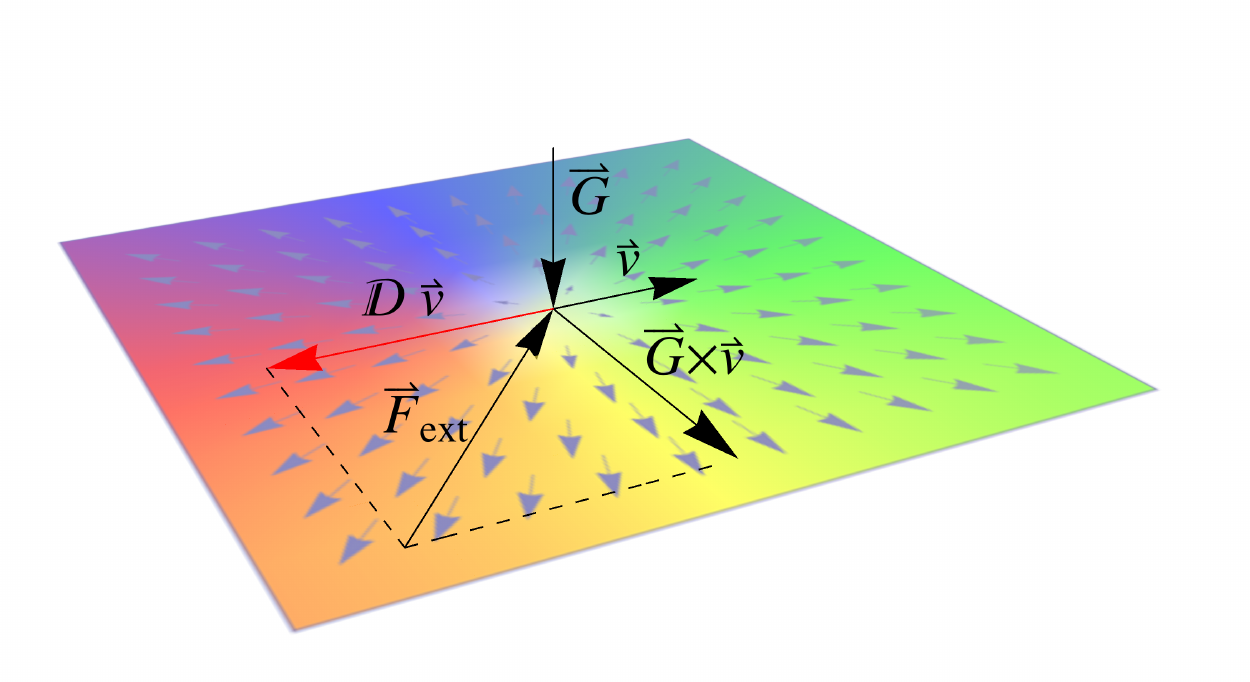}
 \caption{\label{fig:vortexsketch}Sketch of eq.~\eqref{eq:Thiele}, the
   color coding representing the orientation of the spins in the
   $x/y$-plane. The external force is applied intrinsically by the tip
   to keep the steady velocity $v$, which points in
     $x$-direction.  $\vec G \times \vec v$ points in negative
     $y$-direction. Note that only the $x$-component is dissipative,
   the conservative $y$-component counterbalances the gyroscopic
   force.}
\end{figure}
The energy dissipation, occurring in a moved magnetic structure has
been addressed by Thiele \cite{Thiele73}, under the assumption that
the magnetic structure remains invariant when motion with velocity
$\mathbf v$ sets in. Using the LLG equation, Thiele derived in a
continuum approximation
\begin{equation}
  \mathbf F_\mathrm{ext} + \mathbf G \times \mathbf v + \mathcal D \cdot \mathbf v = 0,
\label{eq:Thiele}
\end{equation}
with $\mathbf F_\mathrm{ext}$ representing an external force
initiating and keeping the motion, $\mathbf G$ being the
gyrovector, and $\mathcal D$ a diadic tensor representing the
dissipation. For a magnetisation configuration given in spherical
coordinates ($\mathbf S = (\cos{\varphi} \sin{\theta}, \sin{\varphi}
\sin{\theta}, \cos{\theta}) $) these quantities read
\footnote{Deviating from ref.~\cite{Thiele73}, the factor $1/2$
  appears in eqs.~\eqref{eqref:GD}. This stems from the different
  energy unit, used in this work. While in the (continuum)
  calculations presented in ref.~\cite{Thiele73} the exchange constant
  $A$ is the relevant parameter, here the atomistic exchange constant
  $J = 2 A a$ is of relevance.}
\begin{subequations}%
\begin{align}
  \mathbf G &= -\frac{1}{2} \int d^2r \; \sin{\theta} \; (\nabla \theta \times \nabla \varphi) \;\;\; \mathrm{and} \\
  \mathcal D &= -\frac{\alpha}{2} \int d^2r \; \left (\nabla \theta
    \otimes \nabla \theta + \sin^2{\theta} \; \nabla \varphi \otimes
    \nabla \varphi \right ),
\end{align}%
\label{eqref:GD}%
\end{subequations}%
with $\otimes$ the dyadic product.  Thus, once the
$v{=}0$-configuration is known, the energy
  dissipation occurring in the driven out-of-equilibrium system can
be calculated using the $v{=}0$-result. In the following, I
will derive the friction force emerging from a magnetic vortex. For a
vortex configuration, $\nabla \theta$ points radially from the vortex
core to the tail, while $\nabla \varphi$ circulates tangentially
around the vortex core. The resulting orientation of $\mathbf G$ is
sketched in fig.~\ref{fig:vortexsketch}. The symmetry results in
vanishing off-diagonals and $z$-components of $\mathcal D$. Assuming
cylindrical coordinates in space ($(x,y,z)= (\rho \cos{\phi}, \rho
\sin{\phi},z)$), the dissipation tensor simplifies to
\begin{equation}
  \mathcal D_0 {=} \mathcal D_{xx}{=}\mathcal D_{yy} {=} - \alpha \pi  \int \rho \left (
\left (\frac{\partial \theta(\rho)}{\partial \rho} \right )^2
+\frac{\sin^2{\theta(\rho)}}{\rho^2} \right ) d\rho.
\label{eq:9}
\end{equation}
Thus, in a cylindrically symmetric configuration a dissipative force
acts against the direction of motion $\mathbf v$. This is puzzling at
first sight, as the external force which is applied by the tip on the
substrate contains also a vertical component $F_y$. This component
counterbalances the gyroscopic term $\mathbf G \times \mathbf v$ which
is not dissipative.

In the following, the equilibrium configuration for a continuum
approximation of the present system is derived to illustrate the
assumptions and stress the universality of the results. The impatient
reader may directly proceed to the result, eq.\eqref{eq:c12}.
Equations~\eqref{eq2}-\eqref{eq3} read for cylindrical space
coordinates and spherical spin components (I skip the dependence of
$\theta$ and $\varphi$ on $\mathbf r$ for brevity here) in the
continuum approximation (\textit{cf.}\ ref.~\cite{Takeno1980})
\begin{align}
\mathcal H^c =& \int d^2r \left \{ \frac{J}{2} (\nabla \theta \cdot \nabla \theta +
\sin^2{\theta} \nabla \varphi \cdot \nabla \varphi) - d_z \cos^2{\theta} \right \} \nonumber \\
  +& w \int d^2r \frac{(2 z^2{-}\rho^2) \cos{\theta}  + 3 z \rho \cos{(\phi{-}\varphi) \sin{\theta}}}{(z^2+\rho^2)^{5/2}}.
\label{eq:contHamiltonian}
\end{align}
As I am not interested in the dynamics in this section, I consider the undamped case, $\alpha=0$. Then, the equations of motion for $\theta$ and $\varphi$ read (\textit{cf.}\ ref.~\cite{Takeno1980})
\begin{equation}
\dot{\theta} = \frac{1}{\sin{\theta}} \frac{\partial \mathcal H^c}{\partial \varphi} \;\; \mathrm{and} \;\; \dot{\varphi} = -\frac{1}{\sin{\theta}} \frac{\partial \mathcal H^c}{\partial \theta}.
\end{equation}
Setting $\dot{\theta}\stackrel{!}{=}0$ and
$\dot{\varphi}\stackrel{!}{=}0$ and solving for $\theta$ and $\varphi$
yields the ground state configuration. The resulting differential
equations read
\begin{align}%
  0 &= -2 \cos{\theta} \nabla \theta \cdot \nabla \varphi -
  \sin{\theta} \Delta \varphi + \frac{3 w z \rho
    \sin{(\phi-\varphi)}}{(\rho^2+z^2)^{5/2}}
\label{eq12} \\
0 &= \frac{\Delta \theta}{\sin{\theta}} - (2 d_z + \nabla \varphi
\cdot \nabla \varphi) \cos{\theta} \nonumber \\ &+ w \frac{\rho^2 - 2
  z^2 + 3 \rho z \cos{(\phi-\varphi)} \mathrm{cot}{\theta}}{(\rho^2 +
  z^2)^{5/2}}. \label{eq13}
\end{align}
Because a general solution is not possible, now cylindrical symmetry
with respect to the tip, 
$\varphi(\mathbf r){=}\phi{+}\epsilon$  and $\theta(\mathbf r) {=}
\theta(\rho)$, is
considered. Equation~\eqref{eq12} then directly leads to
$\epsilon{=}0$, as the first two terms vanish and the sine requires
$\varphi{=}\phi$. Rewriting the anisotropy in terms of the
\textit{vortex core length} $\rho_0{=}1/\sqrt{-2 d_z}$ simplifies
eq.~\eqref{eq13} to
\begin{align}
  0 =& \frac{1}{\sin{\theta}} \left (\frac{\partial^2}{ \partial
      \rho^2} + \frac{\partial \theta}{\rho \partial \rho} \right ) -
  \left (\frac{1}{\rho_0^2} + \frac{1}{\rho^2} \right ) \cos{\theta}
  \nonumber \\ & + w \frac{\rho^2 - 2 z^2 + 3 \rho z
    \mathrm{cot}{\theta}}{(\rho^2 + z^2)^{5/2}}.
\label{eq14}
\end{align}
This equation is analytically still not solvable -- even for $w=0$,
which is assumed from now on. However, if the two limiting cases that
the magnetisation points out of the plane in the core region ($\theta
\approx 0$ for $\rho {\ll} \rho_0$) and in plane in the tail region
($\theta \approx \pi/2$ for $\rho {\gg} \rho_0$) are assumed,
eq.~\eqref{eq14} reads
\begin{equation}
\begin{array}{lll}
  0 = \rho^2  \frac{\partial^2 \theta}{\partial \rho^2} + \rho \left( \frac{\partial \theta}{\partial \rho} \right ) + \left ( \frac{\rho^2}{\rho_0^2} - 1 \right ) \theta \; & \mathrm{for} \; \rho {\ll}\rho_0 \\
  0 = \rho^2 \frac{\partial^2 \theta}{\partial \rho^2} + \rho \left( \frac{\partial \theta}{\partial \rho} \right ) + \left ( \frac{\rho^2}{\rho_0^2} + 1 \right ) (\theta{-} \frac{\pi}{2})\; & \mathrm{for} \; \rho {\gg} \rho_0. 
\end{array}
\end{equation}
\begin{figure}
\centering
\includegraphics[width=.72\columnwidth]{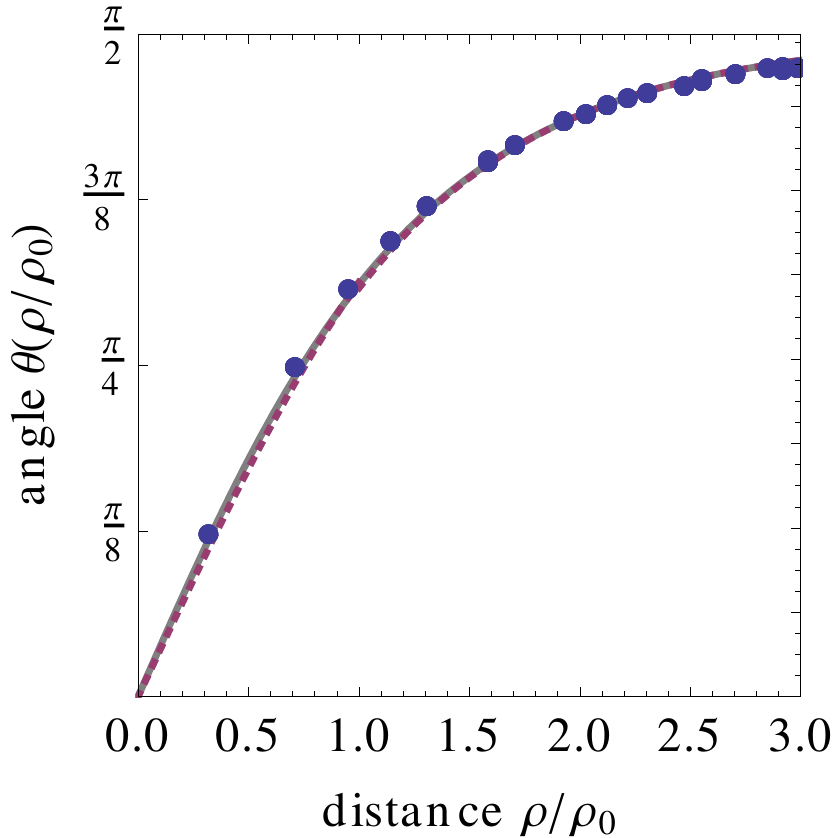}
\caption{\label{fig:solution}Equilibrium solution ($v{=}0$) of the out-of-plane
  magnetisation component $\theta(\rho/\rho_0)$, with $\rho_0 =
  1/\sqrt{-2 d_z}$ being the vortex core radius and
  $\rho=\sqrt{x^2+y^2}$. The black line represent a numerical solution
  of eq.~\eqref{eq14} using $w=0$. The red dotted line represents the
  solution eq.~\eqref{eq:c12}, and the dots the minimisation of
  eq.~\eqref{eq2} on a grid.}
\end{figure}
These are Bessel differential equations and their solutions up to the
integration constants are
\begin{equation}
\theta(\rho) = \left \{ 
\begin{array}{ll} 
c_1 J_1(\rho/\rho_0) & \rho \ll \rho_0 \\
c_2 K_i(\rho/\rho_0) + \pi/2 & \rho \gg \rho_0. 
\end{array} 
\label{eq:c12}\right .
\end{equation}
The constants can be determined from a comparison with numerical
results of eq.~\eqref{eq14} (using $w=0$), $c_1 \approx 2.233$ and
$c_2 \approx -2.081$. The result is plotted in fig.~\ref{fig:solution}.

The $v{=}0$-result can be now plugged into
eq.~\eqref{eq:9}. Because only in the core region the out-of-plane
component of the magnetisation is significant, it is reasonable to
rewrite eq.~\eqref{eq:9},
\begin{equation}
\mathcal D_0 = -\alpha \pi \int \left ( \rho \left (\frac{\partial \theta}{\partial \rho} \right )^2 - \frac{\cos^2{\theta}}{\rho} + \frac{1}{\rho} \right )d\rho. 
\label{eq17}
\end{equation}
\begin{figure}[bt]
\centering
\includegraphics[width=.72\columnwidth]{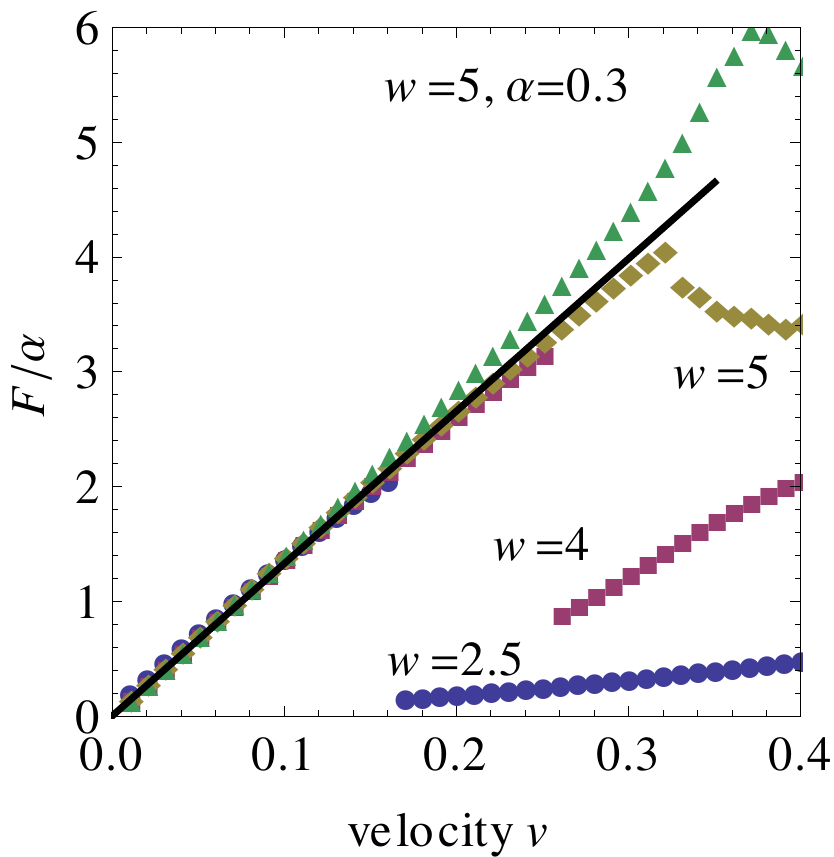}
\caption{\label{fig:Fvsv}Magnetic friction vs.\ velocity for systems
  with a vortex, with a system size $L=48$ and $\alpha=0.5$ (except
  the sample with $\alpha=0.3$). The straight gray line represents
  eq.~\eqref{eq:FlogL}. Above threshold velocities (depending on
  $\alpha$ and $w$), the system makes a transition into a state where
  no longer a vortex is present, \textit{cf.}\ ref.~\cite{Magiera12}.}
\end{figure} 
The first two terms in the integrand of eq.~\eqref{eq17} only
contribute in the core region and quickly converge. The third term
leads to a logarithmic dependence of the dissipation on the system
size and a divergence at the vortex core ($\rho {\rightarrow} 0$). The
latter issue is resolved by introducing a cutoff, which is also
physical as the original lattice model has the lattice constant as a
lower bound for the integration. One may summarise the continuum result
by
\begin{equation}
F = -\mathcal D_0 v=  \alpha \pi v  \log \frac{L}{L_0}, 
\label{eq:FlogL}
\end{equation}
where $L$ is the system size, and $L_0$ a constant which contains the
cutoff at the core, the vortex-core contribution to the integral
(which weakly depends on $d_z$) as well as a geometrical correction as
the continuum is calculated for a disk, whereas our lattice model is a
square. For $d_z=-0.1$ one gets $L_0 \approx 0.7$ from a final
numerical integration of eq.~\eqref{eq17} using eq.~\eqref{eq:c12}.

In fig.~\ref{fig:Fvsv} it is observable that this result, which has
been derived for any vortex dragged by some not specified external
force is also valid for the dipolar tip, as long as $w$ is
  large enough to stabilise a vortex. In practice
a weak dependence of the vortex size $\rho_0$ on $w$ can be
observed. But such minor corrections are always dominated by the
anisotropy, and even the shape of the tip field does not influence
$\rho_0$ significantly. Furthermore the core contribution in
eq.~\eqref{eq17} is always dominated by the tail part, which diverges
with the system size.  One may summarise this section that
  in the quasi-static limit, $v{\rightarrow}0$, the Thiele equation
  provides a good estimate for the energy dissipation.

\section{System-size dependence of friction}
\begin{figure}[bt]
\centering
\includegraphics[width=.72\columnwidth]{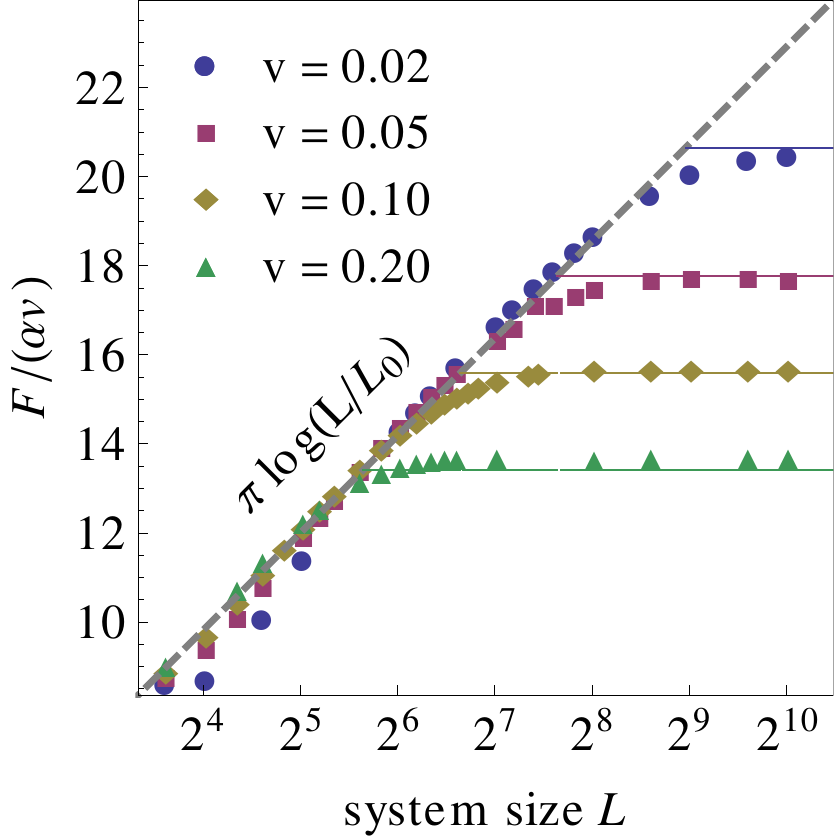}
\caption{\label{fig:FvsL1}Friction force for several velocities vs.\
  the system size for $w=5$ and $\alpha = 0.5$. The horizontal lines
  represent the saturation friction $F_\mathrm{sat}=\pi \alpha v
  \log{(R(\alpha, v)/L_0)}$, assuming eq.~\eqref{eq:vortexsize}.}
\end{figure}
On one hand the logarithmic dependence of the friction force on the
system size leads to a diverging force in the thermodynamic limit. On
the other hand, the quasi-static motion of the ground state is
unphysical in the thermodynamic limit, as the excitations imposed by
the tip have to travel to the system boundary first. This fact has
been already mentioned in ref.~\cite{Huber82}, where the condition
$\alpha =0$ used in that work is made responsible for the logarithmic
divergence in eq.~\eqref{eq:FlogL}. Accordingly, size effects are
important when the system size is increased, and I calculated the
friction force for several velocities and system sizes in computer
simulations.

\begin{figure}[bt]
\centering
\includegraphics[width=.72\columnwidth]{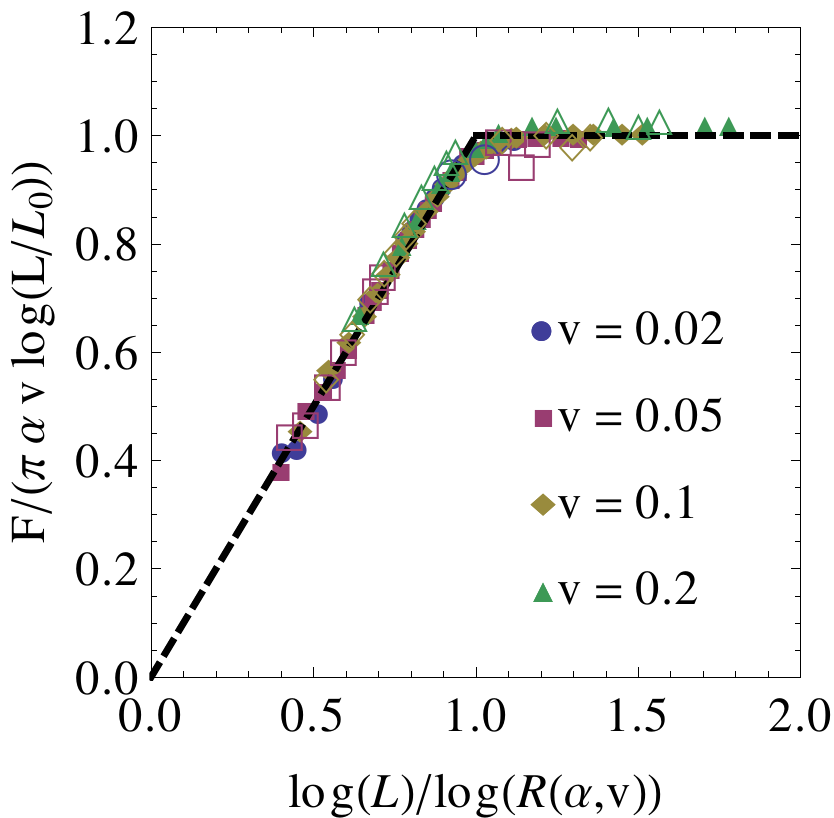}
\caption{\label{fig:crossover}Rescaled friction vs.\ logarithmic and
  rescaled system size, for $\alpha=0.5$ (filled symbols) and $\alpha =
  0.3$ (empty symbols).}
\end{figure}
In fig.~\ref{fig:FvsL1} one finds evidence that eq.~\eqref{eq:Thiele} is
only valid up to a certain system size, which is called an effective
vortex size $R$. Above $R$ the friction force saturates, and from the
value in the large-$L$ limit one may fit the dependence of $R$ on the
dynamic non-equilibrium parameters $v$ and $\alpha$, resulting in
\begin{equation}
R \approx \frac{5 a^2 \omega}{\alpha v}.
\label{eq:vortexsize}
\end{equation}
This can be understood in terms of a macrospin model, introduced in
ref.~\cite{Magiera09a}. If a single Heisenberg spin is driven by an
external field, it tries to follow the field with a lag proportional
to $\alpha v$, where $v$ is the rate of the drive. Approaching
equilibrium (corresponding to a tip at rest,
  $v{\rightarrow}0$), one gets a lag $\alpha v {\rightarrow} 0$,
which corresponds to a vortex with infinite effective size $R$. The
validity of eq.~\eqref{eq:vortexsize} leads to a good data collapse,
\textit{cf.}\ fig.~\ref{fig:crossover}, where both, velocity and
damping constant, have been varied.

\begin{figure}[bt]
\centering
\includegraphics[width=\columnwidth]{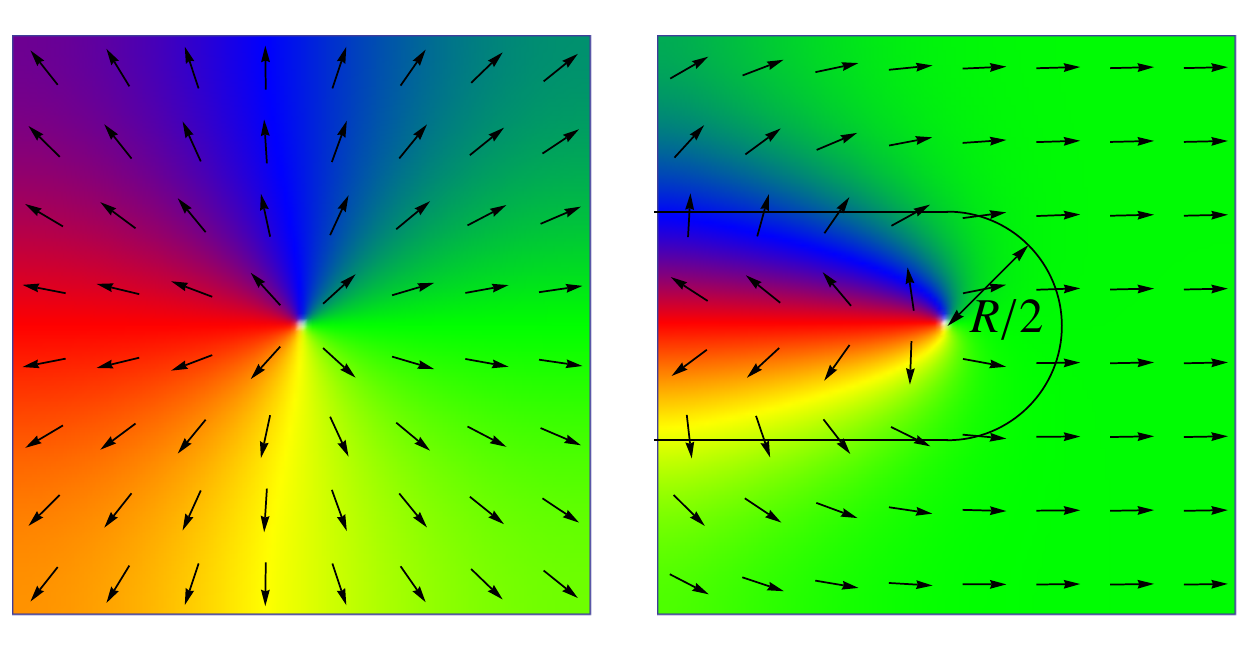}
\caption{\label{fig:snapshots}Snapshots of a system with $L=128$ and
  $\alpha=0.5$ for $v=0.01$ (left, the magnetic friction number is
  $\mathcal M = 16/25$) and $v=0.2$ (right, magnetic friction number
  $\mathcal M = 64/5$). The arrows indicate the local magnetisation
  orientation. Additionally, the effective vortex size is sketched for
  the $v=0.2$ case (it exceeds the system size for the
  $v=0.01$-case).}
\end{figure}

What does a finite effective vortex size mean?  To provide a better
understanding, in fig.~\ref{fig:snapshots} the non-equilibrium steady
state for a system with $L{<}R(\alpha, v{=}0.01)$, now called system
(a), as well as for a system with $L{>}R(\alpha, v{=}0.2)$ (system
(b)) is plotted. Both systems contain exactly one vortex and no
antivortex, with the core directly under the tip, and thus contain the
same vorticity. But while configuration (a) is nearly symmetric with
respect to the $y$-axis, system (b) shows a strong deformation. Spins
at $|y|>R/2$ (the vortex core is at $y=0$) are only little influenced
by the vortex. As a consequence, this part of the system contributes
only a negligibly small $\partial_t \mathbf S_i$-value to the overall
dissipation (\textit{cf.}\ eq.~\eqref{eq:Pdiss}). In other words:
Depending on the parameters $\alpha$ and $v$, the vortex dragged
through the system by the tip has a finite cross-section with a
diameter $R$.

Generalising these observations, I propose a dimensionless
  ``magnetic friction number'', analogous to a magnetic Reynolds
  number
\begin{equation}
\mathcal M = v L \times \frac{\alpha}{\omega a^2}
\label{eq:MR}
\end{equation}
that quantifies the degree of nonlinearity in the response of a driven
spin configuration. If $\mathcal M$ is smaller than unity the Thiele
equation applies yielding the correct dissipation. This statement
should remain valid for structures not containing a vortex, but
\textit{e.g.}\ a domain wall. However, if $\mathcal M$ is greater than
unity, the moved structure is subject to dynamical changes depending
on dynamical parameters $\alpha$ and $v$. Then the detailed
microscopic out-of-equilibrium behaviour and the nonlinearity gains
importance. At the same time, in this limit size effects vanish. For
the case of a vortex configuration this leads to the observed reduced
vortex size $R<L$.

The fraction in eq.~\eqref{eq:MR} is a material specific constant, and
corresponds to about $\mathcal M/(v L) \approx 10^{4}$ s/m$^2$ for
the magnetic transition metals cobalt, iron and nickel. While
nanometre-sized systems should be in the low-$\mathcal M$ regime,
where the equilibrium configuration is relevant and the effects of
finite system size become apparent, fast moving structures in
micron-sized systems make a crossover into the high-$\mathcal M$
regime where non-equilibrium gains importance.

\section{Conclusion}
Energy dissipation for a magnetic vortex, dragged by a dipole tip
through a substrate has been calculated analytically and recovered in
simulations. These results are valid for a vortex driven by any
external force, thus also the drive by a spin-polarised current or an
external magnetic field are possible.  Limitations of the assumptions
which are essential for the analytical result have been discussed. In
a study of size-effects a finite vortex size has been observed which
defines the limit of validity of the analytical result. As the vortex
size is a macroscopic quantity which directly depends on the
microscopic damping constant, the results offer an alternative to
determine the damping constant experimentally or verify the underlying
model, including the verification of the observed dissipation
mechanism.

Based on the observations in the presented system, a magnetic friction
number as a criterion for the validity of the Thiele equation has been
proposed. Its relevance for different system setups, magnetisation
structures as well as \textit{e.g.}\ in the weak damping limit should
be clarified in future studies.
\begin{acknowledgments}
  I am indebted to Dietrich E.\ Wolf for several inspiring discussions
  and a careful reading of the manuscript. I thank also L.\ Brendel
  and A.\ Hucht for several fruitful discussions. This work was
  supported by the German Research Foundation (DFG) through SFB 616
  ``Energy Dissipation at Surfaces''. Computing time by the Neumann
  Institute for Computing (NIC) is gratefully acknowledged.
\end{acknowledgments}

\bibliographystyle{eplbib}
\bibliography{paper}

\end{document}